\documentclass[twocolumn,showpacs,amsmath,amssymb,nofootinbib,aps,prd]{revtex4-1}
\usepackage{dcolumn}
\usepackage{bm}
\usepackage{graphicx}
\usepackage{color}

\usepackage{amsthm}

\newcommand{\be}{\begin{equation}}
\newcommand{\ee}{\end{equation}}
\newcommand{\ba}{\begin{eqnarray}}
\newcommand{\ea}{\end{eqnarray}}
\newcommand{\non}{\nonumber}
\newcommand{\n}[1]{\label{#1}}
\newcommand{\eq}[1]{(\ref{#1})}
\newcommand{\hh}{\, ,\hspace{0.25cm}}
\newcommand{\hhh}{\, ,\hspace{0.5cm}}
\newcommand{\BM}[1]{{\mbox{\boldmath $#1$}}}
\newcommand{\ind}[1]{\mbox{\tiny #1}}
\newcommand{\bi}[1]{\bibitem{#1}}

\begin{document}

\title{Gravitational Faraday and Spin-Hall Effects of Light}
\author{Andrey A. Shoom}
\email{andrey.shoom@aei.mpg.de}
\affiliation{Max Planck Institute for Gravitational Physics (Albert Einstein Institute), Leibniz Universität Hannover, Callinstr. 38, D-30167, Hannover, Germany}


\begin{abstract}

The gravitational Faraday and its dual spin-Hall effects of light arise in space-times of non-zero angular momentum. These effects were studied in stationary, asymptotically flat space-times. Here we study these effects in arbitrary, non-stationary, asymptotically flat space-times. These effects arise due to interaction between light polarisation and space-time angular momentum. As a result of such interaction, the phase velocity of left- and right-handed circularly polarised light becomes different, that results in the gravitational Faraday effect. This difference implies different dynamics of these components, that begin to propagate along different paths\textemdash the gravitational spin-Hall effect of light. Due to this effect, the gravitational field splits a multicomponent beam of unpolarized light and produces polarized gravitational rainbow. The component separation is an accumulative effect observed in long range asymptotics. To study this effect, we construct uniform eikonal expansion and derive dynamical equation describing this effect. To analyse the dynamical equation, we present it in the local space and time decomposition form. The spatial part of the equation presented in the related optical metric is analogous to the dynamical equation of a charged particle moving in magnetic field under influence of the Coriolis force.

\end{abstract}

\maketitle

\section{Introduction}

Gravitational field affects propagation of electromagnetic waves, in particular light, in different ways. For example, electromagnetic radiation emitted by hot accretion disk around a black hole into the external space gets gravitationally redshifted. Light rays passing by a strongly gravitating massive object, e.g. a star or a black hole, get deflected due to the space-time curvature in the vicinity of a massive object. The rays deflection depends on the object's mass and angular momentum. There is also the gravitational Faraday effect, analogical to the magneto-optical Faraday effect\textemdash a rotation of the polarization plane of a linearly-polarized light propagating in a transparent material in the presence of a magnetic field along propagation of the light \cite{LL8}. The gravitational Faraday effect is a rotation of the plane of polarization of an electromagnetic wave propagating in a stationary gravitational field, for example, near a stationary rotating black hole. The study and observation of this effect have quite long history (see, e.g., \cite{Skrotskii,Plebanski,God,GF2,GF3,GF4,GF5,GF6,CariniRuffini,GF7,GF8,GF9,GF10,Ghosh}.) However, despite the clear analysis done in \cite{Plebanski,GF2}, in some works rotation of the polarization vector around direction of light propagation, which is due to its coupling to the space-time angular momentum, is mixed with change of its direction due to deflection of the light rays. 

It is known that in optics there is an effect dual to the Faraday effect\textemdash the optical Magnus effect, that results in the action of light polarization on its trajectory, causing its transverse polarization-dependent displacement. Due to this effect, a linearly polarized light splits into left- and right-handed circularly polarized components propagating along different paths. This splitting effect is known as the spin-Hall effect of light \cite{SH3,SH4,SH5,SH6,SH7}. Spin-Hall effects became quite ubiquitous in modern physics. They are observed in condensed matter \cite{Xiao:2009rm}, optical \cite{Onoda:2004zz}, and high-energy systems \cite{Berard:2004xn}. Classical and quantum spin-Hall optical effects of light are described and analyzed in many works (see, e.g., \cite{SH3,SH4,SH5,SH6,SH7,Bliokh1,Bliokh2,Bliokh3,Bliokh4}.) The underlying nature of these effects is the spin-orbit interaction between the spin of a photon, an electron, or an atom, and its extrinsic angular momentum.

We may expect a dual to the gravitational Faraday effect\textemdash the gravitational spin-Hall effect of light. The gravitational Faraday and spin-Hall optical effects may not be surprising phenomena if one takes a certain point of view on the gravitational field. Namely, one can observe that the source-free Maxwell equations in curved space-time, i.e., in the presence of gravity, can formally be considered as equations in flat space-time in the presence of a bi-anisotropic moving medium whose optical properties are defined by dielectric permittivity and magnetic permeability tensors expressed though the space-time metric tensor components. This point of view was proposed and developed already more than half-century ago by several authors \cite{Skrotskii,Plebanski,Volk}. There is also a dual point of view on propagation of light in the so-called metamaterials, whose optical properties can be synthetized on a subwavelength scale that allows to control propagation of light in a nearly arbitrary way \cite{Shalaev}. In this new field of transformation optics a metric approach can be used to calculate dielectric permittivity and magnetic permeability tensors of a metamaterial \cite{Crudo}. Another example is the gravitational analogue of the linear magnetoelectric effect that was studied in \cite{Gibbons}.

Taking into account the analogy between a gravitational field and a bi-anisotropic moving medium, polarization of light was considered to describe its propagation in a stationary gravitational field. For example, by using the weak field approximation, it was shown that left- and right-handed circularly polarized light propagating near a rotating gravitational body get scattered in a different way \cite{Mashhoon1,Mashhoon2,Mashhoon3,Mashhoon4}. To consider this effect in a strong stationary gravitational field, the so-called modified geometric optics formalism was introduced \cite{Frolov:2011mh}. This formalism was applied to describe scattering of a polarized light propagating in the stationary space-time of a rotating (Kerr) black hole \cite{Frolov:2012zn}. Later this approach was reformulated to some extend \cite{Yoo} and in different context \cite{Sam} in 4-dimensional covariant form. 

There are different approaches have been taken to describe dynamics of polarized light in a curved space-time background (for a review see \cite{Oancea:2019pgm}). For instance, approach based on dynamics of massless spinning particle was proposed and developed in \cite{Duval1,Duval2,Duval3,Duval4} for Riemannian and pseudo-Riemannian manifolds. A semiclassical approach to describe photon dynamics in a curved space-time background based on the Bargmann-Wigner equations was taken in \cite{Static}, and helicity-dependent photon's evolution was predicted for the Schwarzschild space-time. Spin-Hall effect of light for the Schwarzschild spacetime was also predicted in \cite{Oancea:2020khc}. However, in these works the proper orientation and propagation of the basis representing optic axes is not discussed. Thus it remains unclear how to measure properly the evolution of the light polarization along a null ray. As a result, the proposed effect is questionable. Such prediction is also contradictory to the analysis presented in, e.g.,  \cite{Plebanski,Mashhoon1,Frolov:2011mh,Frolov:2012zn,Leite}, where it was shown that left- and right-handed polarization modes evolve differently due to the space-time angular momentum only, and in static space-times there is no distinction between propagation of these modes. In other words, absence of the Gravitational Faraday effect in static space-times implies that these modes evolve in the same fashion. 

To support a possibility of the spin-Hall effect of light in static space-times, one may appeal to the theoretical and experimental studies of the polarization-dependent deflection of light in a smoothly inhomogeneous isotropic medium, where such effect was observed for light propagating through a planar (without torsion) optical fiber (see, e.g., \cite{SH3,Planar}). Note, however, that according to \cite{KNF} (and also the references therein) these polarization-dependent effects observed in planar curved optical fibers are of the higher order. It is also stated in \cite{SH4} that there is no polarization-dependent ray shift can be observed in a planar waveguide, where the Rytov-Vladimirski-Berry phase \cite{Rytov,Vladimirski,Berry,TPhase}, that determines rotation of the polarization plane, vanishes identically. We would also like to note that a gravitational field is analogical to a special kind of bi-anisotropic moving optical medium, such that there is no birefringence and its index of refraction is different for electromagnetic waves propagating in opposite directions \cite{Volk}.\footnote{One may try to describe this phenomenon in the language of Finsler geometry, using properties of the Randers metrics. For a nice review of the Finsler geometry and related problems see \cite{Gibbons:2011ib}.} In other words, a gravitational field is essentially different from a material optical anisotropic medium. Finally, we note that motion of the medium also makes significant contribution to observed optical phenomena and often gives rise to new effects, see, e.g., \cite{Khromykh,VolKis,Privalov,BolStol}. Detailed investigation of analogy between a material optical medium and a gravitational field in context of the related optical effects goes beyond the scope of this paper.    

In this paper, to describe the gravitational Faraday and spin-Hall effects of light, we shall take the modified geometric optics approach \cite{Frolov:2011mh} and extend it to arbitrary non-stationary asymptotically flat space-times of non-zero angular momentum \cite{Ashtekar}. In addition to stationary rotating black holes and stars, such space-times correspond to dynamical gravitational fields due to a gravitational collapse, black holes and neutron stars coalescence, and gravitational waves. They can also represent some cosmological models. Thus, these gravitational optical effects can be widely present. To describe properly these effects, the key property of a stationary space-time, existence of a time-like Killing vector field, was exploited. Here we extend these results by considering a field of {\em static observers}. The key property of such observers is that in their frame, at the spatial infinity $i^0$, the space-time total ADM 3-momentum vanishes. The field of static observers naturally generalises the field of Killing observers and coincides with it in stationary space-times. 

This paper is organized as follows. In Section II we briefly review the laws of (canonical) geometric optics in a curved space-time. In the next Section we study the gravitational Faraday effect of light in arbitrary (non-stationary) space-time. In Section IV we define the field of static observers that can properly detect and measure the gravitational Faraday effect. In Section V we construct the uniform eikonal expansion that takes into account contribution of light polarization to its propagation and present dynamical equations that describe the gravitational spin-Hall effect of light, which is dual to the gravitational Faraday effect. The gravitational spin-Hall effect of light is presented in the local space and time decomposition form in Section VI. The last Section VII contains discussion of the derived results.

Here we shall use geometrized units $c=G=1$ and conventions adopted in the book \cite{MTW}.  

\section{Geometric Optics}

Finding an exact electromagnetic wave solution to the Maxwell equations in a curved space-time background is a formidable problem. Moreover, often such solutions cannot be presented in a closed analytic form. However, for waves that are highly monochromatic over some space-time regions, an asymptotic short-wave (geometric optics) approximation can be used \cite{MTW}. Such approximation allows us to capture basic characteristics of light propagation in a curved space-time background defined by metric $g_{\alpha\beta}$ of the most general form. 

The source-free Maxwell wave equation for the vector potential $A^{\alpha}$ in the Lorenz gauge
\be\n{0a}
A^{\alpha}_{\,\,\,;\alpha}=0\,
\ee
reads
\be\n{0b}
-A^{\alpha;\beta}_{\,\,\,\,\,\,\,\,;\beta}+R^{\alpha}_{\,\,\,\beta}A^{\beta}=0\,,
\ee
 where the semicolon stands for the covariant derivative associated with the space-time metric $g_{\alpha\beta}$ and $R^{\alpha}_{\,\,\,\beta}$ is the 4-dimensional Ricci tensor. The geometric optics approach is based on splitting of the vector potential into a rapidly changing real phase, the {\em eikonal} $\theta$, and a slowly changing complex amplitude in the following way:
\be\n{1}
A^{\alpha}=\Re\{(a^{\alpha}+\varepsilon b^{\alpha}+...)e^{i\theta/\varepsilon}\}\,,
\ee
where $\varepsilon\ll1$ is a dummy expansion parameter that helps to track order of terms: a term with $\varepsilon^n$, for some integer $n$, varies as $(\lambdabar/l_{min})^n$, where $\lambdabar/l_{min}\ll1$. Here $\lambdabar$ is the reduced wavelength (wavelength/$2\pi$) and $l_{min}$ is the minimal of the two characteristic scales\textemdash the curvature radius of the wave front, or the length of a wave packet, and the local curvature radius of the space-time. Substituting the vector potential into the Lorentz gauge condition \eq{0a} and the wave equation \eq{0b} and collecting the leading terms of order $\varepsilon^{-2}$ and $\varepsilon^{-1}$ we derive the fundamental laws of geometric optics: 
\ba
k^{\alpha}k_{\alpha}=0&\hh&k^{\beta}k^{\alpha}_{\,\,\,;\beta}=0\,,\n{2a}\\
k^{\alpha}f_{\alpha}=0&\hh&k^{\beta}f^{\alpha}_{\,\,\,;\beta}=0\,,\n{2b}\\
&&\hspace{-1cm}(a^{2}k^{\alpha})_{;\alpha}=0\,.\n{2c}
\ea
Here $k^{\alpha}=dx^{\alpha}/d\lambda$ is the wave vector metrically related to the gradient $k_{\alpha}\equiv\theta_{;\alpha}$ and tangent to the light ray $\Gamma\!\!:x^{\alpha}=x^{\alpha}(\lambda)$, where $\lambda$ is affine parameter of the ray,  $a\equiv(a^{\alpha}a^{*}_{\alpha})^{1/2}$ is the scalar amplitude, and $f^{\alpha}\equiv a^{\alpha}/a$ is a unit complex polarization vector, such that $f^{\alpha}f_{\alpha}=0$ and $f^{\alpha}f^{*}_{\alpha}=1$. Here and in what follows the superscript $*$ stands for complex conjugation. These laws imply that light rays are the space-time null geodesics \eq{2a}, the polarization vector is orthogonal to the light ray and parallel-propagated along it \eq{2b}, and the vector $a^2k^{\alpha}$ is a conserved current, which defines the adiabatically conserved number of light rays, or in quantum language, the number of photons \eq{2c}. The laws of geometric optics \eq{2a}--\eq{2c} reflect only an approximate picture of light propagation in a curved space-time. In this description polarization of light does not affect its path. 

\section{Gravitational \\ Faraday Effect of Light}

To measure angle of rotation of the polarization plane in the magneto-optical Faraday effect, we have to align properly optic axes of a polarizer and an analyzer. For example, we can align the polarizer and the analyzer at the polarizer's location and then parallel transport the analyzer along the ray trajectory. In a curved space-time this procedure is not so simple.  

Let us present the polarization vector $f^\alpha$ in the following form:\footnote{The polarization vector is defined modulo the wave vector $k^{\alpha}$. This gauge freedom does not affect the results that follow. We shall fix this gauge, as well as the rotation gauge transformation $m^{\alpha}\to m^{\alpha}\exp(i\psi)$, later.} 
\be
f^{\alpha}=e^{i\varphi}m^{\alpha}\,,\n{3}
\ee
where $m^{\alpha}$ is a unit complex vector, such that 
\ba
&&m^{\alpha}m_{\alpha}=0\hhh m^{\alpha}m^{*}_{\alpha}=1\hhh m^{\alpha}k_{\alpha}=0\,,\n{3a}\\
&&m^{\alpha}\equiv\tfrac{1}{\sqrt{2}}(e_1^{\alpha}+i\sigma e_2^{\alpha})\hh m^{*\alpha}\equiv\tfrac{1}{\sqrt{2}}(e_1^{\alpha}-i\sigma e_2^{\alpha})\,,\n{3b}
\ea 
where $e_{1,2}^{\alpha}$ are real orthonormal space-like vectors. The local complex basis $\{m^\alpha, m^{*\alpha}\}$ plays a role of optic axes. To specify the polarization of a given wave we use the parameter $\sigma=\pm1$, with $`+$' for the right- and `$-$' for the left-handed circularly polarized light \cite{MTW}. This definition means that the polarization vector of the (left)right-handed circularly polarized light rotates in the (anti)clockwise direction, when viewed from the source. To define a change in the rotation of the polarization vector along the light ray we introduced {\em polarization phase} $\varphi$.  The polarization phase defines the relative change of the polarization vector along a null ray. For example, in a vacuum and flat space-time $\varphi$ has a constant value in the basis $\{m^{\alpha}, m^{*\alpha}\}$ parallel-transported along the ray. The polarization phase defines an additional angular shift of the polarization vector due to a gravitational field. 

Using the propagation equation for the polarization vector \eq{2b} and the expression \eq{3}, we derive the propagation equation for the polarization phase along the null ray,
\be\n{4}
k^{\alpha}\varphi_{;\alpha}=im^{*}_{\alpha}k^{\beta}m^{\alpha}_{\,\,\,;\beta}\,.
\ee
In a space-time decomposition, spatial part of this expression, corresponding to propagation of the polarization phase along the null ray trajectory, written in momentum parametrisation, is the Rytov-Vladimirski-Berry phase \cite{Rytov,Vladimirski,Berry,TPhase}. This phase of light propagating in a helical optical fiber was experimentally measured and discussed in \cite{Chiao,Tomita}. 

To compute the polarization phase for a given null ray, we have to define a propagation law for $m^{\alpha}$ along the ray. This can be done by an observer-defined local decomposition of the space-time into space and time. This is the so-called space-time threading approach, in contrast to the space-time slicing, which is known as the ADM approach.\footnote{The threading point of view was originally developed by M\o ller, Zel’manov, and Cattaneo, and discussed in detail in \cite{Jantzen,Carini}. It is used in \cite{LL2}.} 

Consider a family of observers filling a 3-dimensional space like a continuous medium. Each of the observers defines the local frame of reference. World lines of these observers form a congruence of integral curves of the timelike future directed unit vector field $u^\alpha=u^\alpha(x^\alpha)$, $u^\alpha u_\alpha=-1$. The local rest space $\Sigma_{u}$ orthogonal to $u^{\alpha}$ is a 3-dimensional subspace of the tangent space defined at every event on an observer's world line. A vector from the tangent space can be projected into the subspace $\Sigma_{u}$ by means of the projection operator $p_{\beta}^{\alpha}=\delta_{\beta}^{\alpha}+u^{\alpha}u_{\beta}$, and $p_{\alpha\beta}=g_{\alpha\beta}+u_{\alpha}u_{\beta}$ defines the induced metric on $\Sigma_{u}$. Applying the projection operator to $k^{\alpha}$ we construct the unit spacelike vector $n^{\alpha}$ that defines the spatial direction of a light ray. Accordingly, we have
\be\n{4a}
k^{\alpha}=\omega(u^{\alpha}+n^{\alpha})\,,
\ee
where $\omega\equiv-k_{\alpha}u^{\alpha}$ is the angular frequency of light measured by the local observer. This decomposition allows us to express propagation of $m^\alpha$ along $k^{\alpha}$ by defining its propagation along the vectors $u^\alpha$ and $n^\alpha$, 
\be\n{4a1}
k^{\beta}m^{\alpha}_{\,\,\,;\beta}=\omega(u^{\beta}m^{\alpha}_{\,\,\,;\beta}+n^{\beta}m^{\alpha}_{\,\,\,;\beta})\,.
\ee
We require that the basis vectors $m^\alpha$ and $m^{*\alpha}$ belong to $\Sigma_{u}$, and thus, according to \eq{3a} and \eq{4a}, are orthogonal to $n^{\alpha}$. Because the polarization vector $f^{\alpha}$ is defined modulo $k^{\alpha}$, this requirement can be fulfilled at some event on the null ray. Then, as it is shown below, this orthogonality condition is preserved along the ray. 

Next we construct a right-handed, observer-adapted orthonormal frame $\{e_0^\alpha,\,e_a^{\alpha},\,a=1,2,3\}$, where $e_0^\alpha\equiv u^{\alpha}$,  $e_{1,2}^{\alpha}$ are defined in \eq{3b}, and $e_3^{\alpha}\equiv n^{\alpha}$. For such frame we have $\varepsilon_{\alpha\beta\gamma\delta}e_0^{\alpha}e_1^{\beta}e_2^{\gamma}e_3^{\delta}=+1$, where $\varepsilon_{\alpha\beta\gamma\delta}$ is the Levi-Civita (pseudo) tensor. Using \eq{4a}, this gives 
\be\n{4b}
\varepsilon_{\alpha\beta\gamma\delta}u^{\alpha}k^{\beta}m^{*\gamma}m^{\delta}=i\sigma\omega\,,
\ee
that implies
\be\n{5}
\varepsilon_{\alpha\beta\gamma\delta}m^{*\gamma}m^{\delta}=\frac{i\sigma}{\omega}
(k_{\alpha}u_{\beta}-u_{\alpha}k_{\beta})\,.
\ee
We shall also need the following property of the Levi-Civita tensor:
\be\n{5a}
\varepsilon_{\mu\nu\gamma\delta}\varepsilon^{\mu\nu\alpha\beta}=-2(\delta_{\gamma}^{\alpha}\delta_{\delta}^{\beta}-\delta_{\delta}^{\alpha}\delta_{\gamma}^{\beta})\,,
\ee
where $\delta^{\alpha}_{\beta}$ is the 4-dimensional Kronecker tensor.

To measure properly the polarization phase, we require first that the basis $\{m^\alpha, m^{*\alpha}\}$ does not rotate with respect to a reference basis fixed at the spatial infinity, when it is spatially transported along a ray trajectory\footnote{For description of optical measurements in curved space-time see, e.g., \cite{Synge}.}, and second that its initial orientation does not change when it is transported along the observer congruence. The first requirement is ensured by the vanishing spatial Fermi-Walker derivative of $m^{a}$ along $n^{a}$,
\be\n{6}
\nabla^{\ind{FW}}_{n}m^{a}\equiv n^{b}m^{a}_{\,\,\,|b}-(a^{a}n^{b}-a^{b}n^{a})m_{b}=0\,.
\ee 
Here $a^{a}=n^{b}n^{a}_{\,\,\,|b}$ and the stroke $|$ stands for the covariant derivative associated with the spatial metric $p_{ab}=e_a^{\alpha}e_b^{\beta}p_{\alpha\beta}$, such that $p_{ab|c}=0$. This derivative is related to the covariant derivative associated with the space-time metric $g_{\alpha\beta}$ as follows: $e_a^{\alpha}n^bm^a_{\,\,\,|b}=p_{\gamma}^{\alpha}n^{\beta}m^{\gamma}_{\,\,\,;\beta}$. The orthogonality condition $m^{\alpha}n_{\alpha}=m^{a}n_{a}=0$ is preserved by the Fermi-Walker derivative. To fulfill the second requirement, we impose that the basis $\{m^\alpha, m^{*\alpha}\}$ is co-rotating with the congruence. This implies that the basis has no relative temporal rotation with respect to nearby observers (and therefore with respect to the reference basis fixed at the spatial infinity.) This requirement is ensured by the vanishing temporal co-rotating Fermi-Walker derivative of $m^{\alpha}$ along $u^{\alpha}$ (see, e.g., \cite{Jantzen}),
\be\n{7}
\nabla^{\ind{CFW}}_{u}m^{\alpha}\equiv u^{\beta}m^{\alpha}_{\,\,\,;\beta}+(w^{\alpha}u^{\beta}-w^{\beta}u^{\alpha})m_{\beta}-\omega^{\alpha}_{\,\,\,\beta}m^{\beta}=0\,.
\ee 
Here $w^{\alpha}=u^{\beta}u^{\alpha}_{\,\,\,;\beta}$ is 4-acceleration and $\omega_{\alpha\beta}=p_{\alpha}^{\gamma}p_{\beta}^{\delta}u_{[\gamma;\delta]}$ is the vorticity tensor. The first three terms represent the temporal Fermi-Walker derivative. The orthogonality condition $m^{\alpha}u_{\alpha}=0$ is preserved by the co-rotating Fermi-Walker derivative. The conditions \eq{6} and \eq{7} fix the scalar function $\psi=\psi(x^{\alpha})$ in the gauge transformation $m^{\alpha}\to m^{\alpha}\exp(i\psi)$. 
 
Using the decomposition of the null vector $k^{\alpha}$ \eq{4a}, the expressions \eq{4a1}, \eq{5}, \eq{5a}, and the transport laws \eq{6} and \eq{7}, we can calculate the right-hand side of \eq{4} as follows,
\be
im^{*}_{\alpha}k^{\beta}m^{\alpha}_{\,\,\,;\beta}=i\omega m^{*\alpha}\omega_{\alpha\beta}m^{\beta}=\sigma \omega^{\alpha}k_{\alpha}\,,\n{8a}
\ee
where
\be
\omega^{\alpha}=\tfrac{1}{2}\varepsilon^{\alpha\beta\gamma\delta}u_{\beta}\omega_{\gamma\delta}=\tfrac{1}{2}\varepsilon^{\alpha\beta\gamma\delta}u_{\beta}u_{\gamma;\delta}\,,\n{8b}
\ee
is the vorticity of the observers congruence. 

By using this result, we can now compute the polarization phase $\varphi$ for a given null ray $\Gamma\!\!:x^{\alpha}=x^{\alpha}(\lambda)$,
\be\n{9}
\varphi=\sigma\int_{\Gamma}\omega^{\alpha}k_{\alpha}d\lambda=\sigma\int_{\Gamma}
\omega_{\alpha}dx^{\alpha}\,.
\ee 

Finally, we can consider a linearly-polarized light, viewed as a superposition of its left- and right-handed circularly polarized components. The linear polarization real unit vector $f^{\alpha}_{L}=(f^{\alpha}+f^{*\alpha})/\sqrt{2}$ rotates with respect to the basis $\{m^{\alpha}, m^{*\alpha}\}$ and the angle of rotation $\varphi_{L}$ measured along the light ray ${\Gamma}$ is 
\be\n{10}
\varphi_{L}=\int_{\Gamma}\omega_{\alpha}dx^{\alpha}\,.
\ee
This rotation is known as the {\em gravitational Faraday effect of light.}

\section{Field of Observers}

So far we have not specified the field of observers $u^{\alpha}$. As it follows from the expressions \eq{9} and \eq{10}, the gravitational Faraday rotation depends on vorticity of the observers congruence. For example, freely falling (inertial) observers do not feel the gravitational field and their congruence has zero vorticity. The same situation happens for the zero angular momentum observers. Thus, such observers do not detect the gravitational Faraday rotation. Alternatively, we can consider a congruence of arbitrarily moving (non-inertial) observers whose congruence has non-zero vorticity. Such observers would claim to detect the gravitational Faraday rotation in a flat space-time. {\em What kind of observers one has to consider in order to measure properly the gravitational Faraday effect?} 

The polarization-dependent gravitational optical effects were studied in stationary space-times. Such space-times posses timelike Killing vector field $\xi^\alpha_{(t)}$, where $t$ is the Killing time, a parameter  along Killing vector field orbits. Naturally, in such space-times the field of Killing observers was taken, $u^\alpha\propto\xi^\alpha_{(t)}$. Here we consider asymptotically flat non-stationary space-times (see, e.g., \cite{Wald}). Such space-times do not possess timelike Killing vector field. In this case, the best one can do is to take an inertial frame in the asymptotically flat region and to construct connected to the frame Cartesian coordinate latticework. Such a latticework is assumed to be absolutely rigid and extends to other regions of the space-time as far as possible.\footnote{Note, however, that such a latticework cannot be extended into certain space-time regions, for example into a rotating black hole's ergosphere or into a black hole interior.} We place identical clocks in every point of the latticework and synchronise them modulo the redshift factor, i.e., (proper time at some point on the latticework) = (redshift factor at that point) $\times$ (proper time on the latticework at the asymptotically flat region). This construction represents a field of observers that are situated at every point of the latticework, i.e. they have fixed spatial coordinate position defined by asymptotically Cartesian coordinates: $(x^i=const,\, i=1,2,3)$. Accordingly, in these coordinates the observers vector field is
\be\n{11}
u^\alpha=\frac{1}{\sqrt{h}}\delta^\alpha_0\,.
\ee
Here $h>0$ is the squared redshift factor and $x^0=t$ is the timelike coordinate that measures proper time of observers sitting on the latticework in the asymptotically flat region. Using timelike threading approach \cite{LL2} we can present the space-time metric in the coordinates $(x^0=t, x^i)$ in the following form:
\be\n{12}
ds^{2}=-h(dt-g_{i}dx^{i})^{2}+h\gamma_{ij}dx^{i}dx^{j}\,.
\ee
Here $h\gamma_{ij}$ is the 3-dimensional metric that defines spatial distance and the metric functions $h$, $g_{i}$, and $\gamma_{ij}$ depend on $t$ and $x^{i}$. Accordingly, the covariant form of the observers field reads
\be\n{13}
u_{\alpha}=-\sqrt{h}\left(\delta^{0}_{\alpha}-g_i\delta^i_{\alpha}\right)\,.
\ee
Let us now compute the vorticity \eq{8b} of the observers field \eq{13} in the metric \eq{12},
\be\n{14}
\omega^\alpha=\frac{1}{2h}\left\{(\BM{g}, \text{curl}\BM{g})\delta^{\alpha}_{0}+([\BM{g}\times\BM{g}^{\flat}_{,t}]^i+(\text{curl}\BM{g})^i)\delta^\alpha_i\right\}\,.
\ee
Here $\BM{g}^{\flat}$ is the covariant form $g_{i}$ of the vector $\BM{g}$ living in a 3-dimensional space endowed with the metric $\gamma_{ij}$, 
\ba\n{14a}
(\BM{a}, \BM{b})&=&a^{i}b^{j}\gamma_{ij}\hhh[\BM{a}\times\BM{b}]_{i}=e_{ijk}a^{j}b^{k}\,,\\
(\text{curl}\BM{g})^{i}&=&e^{ijk}g_{k,j}\hhh e_{ijk}=\sqrt{\gamma}\epsilon_{ijk}\hhh e^{ijk}=\frac{\epsilon^{ijk}}{\sqrt{\gamma}}\,,\non
\ea
where $\gamma=\text{det}(\gamma_{ij})$, $\epsilon_{123}=\epsilon^{123}=1$ is the Levi-Civita symbol, and the indices are raised and lowered by $\gamma_{ij}$ in the usual way. The expressions $(...)_{,t}$ and $(...)_{,i}$ mean partial derivatives of $(...)$ with respect to $t$ and $x^{i}$. Using \eq{14} we can calculate the gravitational Faraday rotation \eq{10}. 

Note, however, that the observers field $u^{\alpha}$ is not unique. One can consider another field of observers $\tilde{u}^{\alpha'}$ that have fixed spatial coordinate position on the related latticework $(x^{i'},\, i'=1',2',3')$. This new lattice work and the proper time $x^{0'}=t'$ of such observers located at the asymptotically flat infinity are related to the former ones by the Lorentz transformation, $x^{\alpha}=\Lambda^{\alpha}_{\,\,\,\beta'}x^{\beta'}$ (see, e.g., \cite{MTW}, p.~69). The above expressions \eq{11}--\eq{13} have the same form in the primed frame. To understand how the gravitational Faraday rotation depends on the observers field, one has to find how the corresponding vorticity expressions are related to each other. To do it, we first derive a relation between the observers vector fields $u^{\alpha}$ and $\tilde{u}^{\alpha'}$ in the frame $x^{\alpha}$, 
\be\n{15}
\tilde{u}^\alpha=\frac{\gamma_g}{1-(\BM{g}, \BM{v})}\left\{u^\alpha+\frac{v^i}{\sqrt{h}}\delta^\alpha_i\right\}\,.
\ee
Here
\be\n{16}
\gamma_g\equiv\frac{1}{\sqrt{1-(\BM{v}_g, \BM{v}_g)}}\hhh \BM{v}_g\equiv\frac{\BM{v}}{1-(\BM{v}, \BM{g})}\,,
\ee
where $\BM{v}$ is the 3-velocity with the constant contravariant components $(v^i=const,\, i=1,2,3)$, which are parameters of the Lorentz transformation.\footnote{In a curved space-time region $v_{i}=\gamma_{ij}v^{j}\ne const.$} Accordingly, the covariant form of the new observers field reads
\be\n{17}
\tilde{u}_{\alpha}=-\gamma_g\sqrt{h}\left(\delta^{0}_{\alpha}-\tilde{g}_i\delta^i_{\alpha}\right)\,,
\ee
where
\be\n{18}
\BM{\tilde{g}}\equiv\BM{g}+\BM{v}_g\,.
\ee
Now we can compute the vorticity \eq{8b} of the observers field \eq{17} in the metric \eq{12},
\be\n{19}
\tilde{\omega}^\alpha=\frac{\gamma^2_g}{2h}\left\{(\BM{\tilde{g}}, \text{curl}\BM{\tilde{g}})\delta^{\alpha}_{0}+([\BM{\tilde{g}}\times\BM{\tilde{g}}^{\flat}_{,t}]^i+(\text{curl}\BM{\tilde{g}})^i)\delta^\alpha_i\right\}\,.
\ee
Thus, we have $\tilde{\omega}^{\alpha}\ne\omega^{\alpha}$ and according to the expression \eq{9}, the gravitational Faraday effect is observer-dependent. In particular, in a static space-time, such that in the frame $x^{\alpha}$ we have $g_{i}=0$, one can find an observers field of non-zero vorticity. Such observers would claim to detect the gravitational Faraday effect proportional to $v^{i}$, i.e. to the parameters of the Lorentz transformation. In the next section we study the gravitational spin-Hall effect of light, which is dual to the gravitational Faraday effect. This dual effect is, in turn, would also be proportional to $v^{i}$. In particular, for $v^{i}=0$ both the effects vanish. This phenomenon is analogical to the relativistic Hall effect resulting in a transverse shift of the relativistic center of inertia of a dynamical system \cite{Bliokh5}. The shift is proportional to the intrinsic angular momentum of the system and to the velocity $v^{i}$ of the relativistic frame, which is moving with respect to the rest frame of the system. The key issue behind the relativistic Hall effect is that components of the 3-dimensional vector of the relativistic center of inertia are not spatial components of a 4-dimensional vector. Thus, they do not transform in covariant way, that makes the location of the relativistic center of inertia frame-dependent  \cite{LL2}.  In our case, the analogical vector is the 3-dimensional vector $\BM{g}$, which transforms according to \eq{18}.

Thus, in the case of a non-static gravitational field $(g_{i}\ne0)$, the general field of observers \eq{15} would detect the gravitational Faraday rotation due to both the gravitational field and their own motion, i.e. via the values of $v^{i}$. This disadvantage is naturally resolved in stationary space-times by selecting the preferred field of Killing observers. Our goal is to find a field of observers that is analogical to the Killing observers in an arbitrary space-time. In other words, we have to fix the kinematic gauge freedom $v^{i}$, that brings us back to the question raised in the first paragraph of this section. 

To answer the question, we note first that an asymptotically flat stationary space-time has vanishing total 3-momentum, as defined with respect to the observers field that coincides with the field of Killing observers. We can take this property as the property that allows us to fix the kinematic gauge and thus to define the observers field in non-stationary space-times. To begin, we recall that asymptotic flatness structure allows the space-time energy-momentum 4-vector $P^{\alpha}=(E,P^i)$ to be well-defined at the spatial infinity $i^0$ as follows (for more details see \cite{Wald}, Ch.~11 and the references therein):  Here we deal with globally hyperbolic space-times. A globally hyperbolic space-time can be foliated by Cauchy hypersurfaces $\Sigma_{t}$ parametrized by a global time function $t$. Consider a unit, time-like, future-directed, vector field $N_{\alpha}\propto t_{;\alpha}$. Then, the space-time metric $g_{\alpha\beta}$ induces a 3-dimensional spatial metric 
\be\n{20}
H_{\alpha\beta}=g_{\alpha\beta}+N_{\alpha}N_{\beta}
\ee
on each $\Sigma_{t}$. Let $\Sigma_{t}$ be such that this metric at $i^0$ in the asymptotically Cartesian coordinates $(x^i,\, i=1,2,3)$ has the form $\delta_{ij}+{\cal O}(1/r)$, where $\delta_{ij}$ is the 3-dimensional Kronecker tensor and $r=\sqrt{x^ix_i}$. Then, the space-time total energy $E$ and 3-momentum $P_{i}$ are defined as follows: 
\ba
E&\equiv&\frac{1}{16\pi}\lim_{r\to\infty}\oint_{\cal S}\left(H_{ij,i}-H_{ii,j}\right)S^{j}dA\,,\n{21a}\\
P_{i}&\equiv&\frac{1}{8\pi}\lim_{r\to\infty}\oint_{\cal S}\left(K_{ij}S^{j}-K^{j}_{\,\,j}S_{i}\right)dA\,,\n{21b}
\ea
where summation over repeated indices is assumed. The integrals are taken over a 2-sphere ${\cal S}\!:r=const$, $S^{i}$ is a unit, outward-directed, space-like vector orthogonal to ${\cal S}$, $dA$ is the area element on ${\cal S}$, which in the limit $r\to\infty$ and in spherical coordinates $(r, \theta, \phi)$ takes the form $dA=r^2\sin^2\theta d\theta d\phi$, and
\be\n{22}
K_{ij}\equiv\frac{1}{2}({\cal L}_{\BM{\scriptscriptstyle N}}\BM{H})_{ij}=\frac{1}{2}(N^{k}H_{ij,k}+H_{kj}N^{k}_{\,\,,i}+H_{ik}N^{k}_{\,\,,j})\,,
\ee
is the hypersurface extrinsic curvature. In this construction, the so-called ADM energy-momentum 4-vector
\be\n{23}
P_{\alpha}=-EN_{\alpha}+P_{i}\delta_{\alpha}^i
\ee
is independent of the choice of $\Sigma_{t}$. As a result, the space-time total energy $E$ and total 3-momentum $\BM{P}$ depend only on the asymptotic behaviour of a spacelike hypersurface $\Sigma_{t}$ at $i^0$ and transform properly under Lorentz boost, i.e., as the components of a 4-vector.  Thus, by an appropriate choice of the $t$ function, or, in other words, taking a proper boost at the asymptotic spatial infinity $i^0$, one can make the space-time total 3-momentum $\BM{P}$ vanish. This choice of $t$ fixes the kinematic gauge and defines the corresponding field of observers, that we shall call {\em static observers}. The field of static observers naturally generalises the field of Killing observers, which is hypersurface-orthogonal at asymptotic infinity. Static observers coincide with Killing observers in stationary space-times. {\em In what follows, to discuss the gravitational Faraday effect and its dual gravitational spin-Hall effect, we shall always consider the field of static observers.}  According to the conditions \eq{6} and \eq{7}, these observers possess unidirectional basis $\{m^\alpha, m^{*\alpha}\}$, adjusted to a reference basis fixed at the spatial infinity, that allows them to measure properly these optical effects.

\section{Gravitational \\ Spin-Hall Effect of Light}

As we already noted, in the geometric optics approach polarization does not affect light rays. A similar situation occurs when one applies the WKB method to the Dirac equation: electric and magnetic particle's moments and spin do not affect its trajectory \cite{Pauli,RubKel}. However, the WKB expansion is not uniformly valid in its domain. At finite fixed distances from inhomogeneous field regions, effects of the particle's moments on its trajectory are of order $\hbar$ and they vanish in the classical limit $\hbar\to0$. In this case, the WKB method gives correct result. But for distances of order $\hbar^{-1}$, the effects become of order unity and do not vanish in the limit $\hbar\to0$. In this case, the WKB method fails. As it was explained in \cite{RubKel}, to obtain an expansion which is uniformly valid everywhere, including the neighbourhood of infinity, one has to include effects of the particle's moments and spin on its trajectory \cite{Schiller1,Schiller2,KNF}. Analogously, to have an eikonal expansion uniformly valid everywhere, one has to take into account contribution of internal degrees of freedom (polarization) to propagation of light \cite{Frolov:2011mh}. Such a contribution is of order $\varepsilon$ for short distances of propagation. However, it accumulates along light ray trajectory and for sufficiently large distances (of order $\varepsilon^{-1}$) it becomes of order $\varepsilon^0$. 

To construct such an expansion we have to include polarization phase into the eikonal.  As we already found, in the geometric optics approximation, 
\be\n{24_0}
A^{\alpha}\approx\Re\{a^{\alpha}e^{i\theta/\varepsilon}\}\hhh a^{\alpha}=a\,m^{\alpha}e^{i\varphi}\,.
\ee
Here the polarization phase $\varphi$ changes along light ray trajectory in accordance with \eq{9}, but this change does not affect the trajectory. Our goal is to modify the light trajectory in accordance with the polarization phase change. To anticipate the $\exp(i\varphi)$ term, let us rewrite this expression in the following form:
\be\n{24}
A^{\alpha}\approx\Re\{a^{\alpha}e^{i\varphi}e^{i(\theta-\varepsilon\varphi)/\varepsilon}\}\,.
\ee
This form suggests us to define the {\em combined eikonal}:
\be\n{25a}
\tilde{\theta}\equiv\theta-\varepsilon\tilde{\varphi}\,,
\ee
where $\tilde{\varphi}$ is the modified polarization phase, 
\be\n{25b}
\tilde{\varphi}\equiv\varphi + \psi\,,
\ee
that corresponds to the combined eikonal $\tilde{\theta}$, and $\psi$ is order $\varepsilon$ contribution to the polarization phase $\varphi$ due to the same order modification in the eikonal \eq{25a}. Note that we could alternatively take the opposite signs in front of $\tilde{\varphi}$ and $\psi$. However, a deeper inspection of the expressions \eq{24_0} and \eq{24} reveals that in \eq{24} the polarization phase $\varphi$ is separated from the amplitude $a^{\alpha}$ with the proper sign, while to keep the vector potential $A^{\alpha}$ equal to its original form \eq{24_0}, it is also subtracted from the eikonal $\theta$. Our choice ensures that the further calculations bring us to the right expression \eq{39}, which implies that the contribution to the polarization phase, $\psi$, is indeed of order $\varepsilon$. 

With these modifications the expression \eq{24} reads:
\be\n{26}
A^{\alpha}\approx\Re\{\tilde{a}^{\alpha}e^{i\tilde{\varphi}}e^{i\tilde{\theta}/\varepsilon}\}\,.
\ee
Here the amplitude $\tilde{a}^{\alpha}$ corresponds to the modified eikonal \eq{25a}. Substituting this approximation into the Lorenz gauge condition \eq{0a} and the wave equation \eq{0b} and holding $\varepsilon$ order terms within $\tilde{\theta}$ and $\tilde{\varphi}$ we derive the following leading order equations:
\be\n{27a}
\mathtt{k}^{\alpha}\mathtt{k}_{\alpha}=0\hhh \tilde{a}_{\alpha}\mathtt{k}^{\alpha}=0\,.
\ee
They imply that the wave vector 
\be\n{27b}
\mathtt{k}_{\alpha}\equiv\tilde{\theta}_{;\alpha}=\theta_{;\alpha}-\varepsilon\tilde{\varphi}_{;\alpha}\,
\ee
is null and electromagnetic waves are transverse. These conditions hold along null ray defined by the wave vector. The term $\tilde{\varphi}_{;\alpha}$ is the gradient of the polarization component of the combined eikonal. For a particular null ray $\tilde{\Gamma}\!\!:\tilde{x}^{\alpha}=\tilde{x}^{\alpha}(\lambda)$, with the wave vector $\mathtt{k}^{\alpha}=d\tilde{x}^{\alpha}/d\lambda$, the polarization phase is [cf. \eq{9}],
\be\n{28}
\tilde{\varphi}=\sigma\int_{\tilde{\Gamma}}\omega^{\alpha}\mathtt{k}_{\alpha}d\lambda=\sigma\int_{\tilde{\Gamma}}
\omega_{\alpha}dx^{\alpha}\,.
\ee 
According to this expression, the gradient of the polarization component of the wave front is $\tilde{\varphi}_{;\alpha}=\sigma\omega_{\alpha}$.

To construct the propagation equation for the wave vector, we use the same method which is used in the geometric optics (see \cite{MTW}, p. 576). Namely, taking the covariant derivative of the first expression in \eq{27a} we derive $\mathtt{k}^{\beta}\mathtt{k}_{\beta;\alpha}=0$, while the expression \eq{27b} and the equality $\theta_{;\beta\alpha}=\theta_{;\alpha\beta}$ give  
\be\n{29}
\mathtt{k}_{\beta;\alpha}=\mathtt{k}_{\alpha;\beta}-\varepsilon\sigma\Phi_{\alpha\beta}\,,
\ee
where we defined
\be\n{30}
\Phi_{\alpha\beta}\equiv\omega_{\beta;\alpha}-\omega_{\alpha;\beta}\,.
\ee
The expression \eq{29} gives us the null rays equation
\be\n{31}
\mathtt{k}^{\beta}\mathtt{k}^{\alpha}_{\,\,\,;\beta}=\varepsilon\sigma\,\Phi^{\alpha}_{\,\,\,\beta}\mathtt{k}^{\beta}\,.
\ee
Note that this equation looks like the Lorentz force law.\footnote{Equation \eq{31} can formally be derived by taking the ultrarelativistic limit $m\to0$ of the Lorentz force law, where $m$ is the rest mass of a charged particle and its proper time is defined as $\tau\equiv m\lambda$ \cite{Frolov:2012zn}. In this analogy the product of the particle's charge and the electromagnetic field tensor $F_{\alpha\beta}$ corresponds to $\varepsilon\sigma\,\Phi_{\alpha\beta}$ and $\omega_{\alpha}$ plays the role of an electromagnetic vector potential. Note that in analogy with the action for a charged particle moving in an electromagnetic field, equation \eq{31} can also be derived from the generalized Fermat's principle \cite{Perlick1,Perlick2,Fermat,Frolov}.}

Now we shall construct propagation equation for the amplitude $\tilde{a}^{\alpha}$. We begin with the transversality condition [the second expression in \eq{27a}]. This condition should hold along a null ray, $\mathtt{k}^{\beta}(\tilde{a}_{\alpha}\mathtt{k}^{\alpha})_{;\beta}=0$. Expanding this expression and using \eq{31} we derive
\be\n{32}
\mathtt{k}^{\alpha}\mathtt{k}^{\beta}\tilde{a}_{\alpha;\beta}=\varepsilon\sigma\Phi_{\alpha\beta}\mathtt{k}^{\alpha}\tilde{a}^{\beta}\,.
\ee
The right-hand side of this expression is of order $\varepsilon$. The corresponding order complement comes from the subleading order $\varepsilon$ wave equation \eq{0b},
\be\n{33}
\mathtt{k}^{\beta}\tilde{a}_{\alpha;\beta}=-\frac{1}{2}\tilde{a}_{\alpha}\mathtt{k}^{\beta}_{\,\,\,;\beta}-i\tilde{a}_{\alpha}\mathtt{k}^{\beta}\tilde{\varphi}_{;\beta}\,.
\ee
Combining these expressions we derive the propagation equation for the complex amplitude $\tilde{a}^{\alpha}$,
\be\n{34}
\mathtt{k}^{\beta}\tilde{a}^{\alpha}_{\,\,\,;\beta}=\varepsilon\sigma\Phi^{\alpha}_{\,\,\,\beta}\tilde{a}^{\beta}-\frac{1}{2}\tilde{a}^{\alpha}\mathtt{k}^{\beta}_{\,\,\,;\beta}-i\tilde{a}^{\alpha}\mathtt{k}^{\beta}\tilde{\varphi}_{;\beta}\,.
\ee
Now we multiply this equation by the complex conjugate amplitude $\tilde{a}^{*\alpha}$ and add to it its complex conjugate form multiplied by $\tilde{a}^{\alpha}$. As a result, we derive
\be\n{35}
(\tilde{a}^2\mathtt{k}^{\alpha})_{;\alpha}=0\,,
\ee
where $\tilde{a}\equiv(\tilde{a}^{\alpha}\tilde{a}_{*\alpha})^{1/2}$ is the scalar amplitude. The derived equation implies adiabatic conservation of photons propagating along null curves defined by \eq{31}. To complete the construction of equations corresponding to the combined eikonal \eq{25a} we introduce the unit complex polarisation vector $\mathtt{f}^{\alpha}\equiv\tilde{a}^{\alpha}/\tilde{a}$, such that $\mathtt{f}^{\alpha}\mathtt{f}_{\alpha}=0$ and $\mathtt{f}^{\alpha}\mathtt{f}^{*}_{\alpha}=1$. Then, the second expression in \eq{27a} and the expression \eq{34} give
\ba
&&\mathtt{k}^{\alpha}\mathtt{f}_{\alpha}=0\,,\n{36a}\\
\mathtt{k}^{\beta}\mathtt{f}^{\alpha}_{\,\,\,;\beta}&=&\varepsilon\sigma\,\Phi^{\alpha}_{\,\,\,\beta}\mathtt{f}^{\beta}-i\mathtt{f}^{\alpha}\mathtt{k}^{\beta}\tilde{\varphi}_{;\beta}\,. \n{36b}
\ea
The expressions \eq{27a}, \eq{31}, \eq{35}, \eq{36a}, and \eq{36b} represent the modified geometric optics corresponding to the combined eikonal \eq{25a}. The last step is to show that in the limit $\varepsilon\to0$ polarization phase $\psi$ does not change along the modified null rays. To begin with we present the polarisation vector $\mathtt{f}^{\alpha}$ in the form [cf. \eq{3}]  
\be\n{37}
\mathtt{f}^{\alpha}=e^{i\psi}\mathtt{m}^{\alpha}\,.
\ee
Here $\psi$ is measured with respect to the complex basis $\{\mathtt{m}^\alpha, \mathtt{m}^{*\alpha}\}$, which is analogical to the previously considered basis $\{m^\alpha, \bar{m}^{\alpha}\}$ (see Sec.~III). Accordingly, one can repeat the steps in Sec.~III and derive [cf. \eq{8a}] 
\be\n{38}
i\,\mathtt{m}_{*\alpha}\mathtt{k}^{\beta}\mathtt{m}^{\alpha}_{\,\,\,;\beta}=\mathtt{k}^{\alpha}\varphi_{;\alpha}\,.
\ee
Then, from the relations \eq{25b}, \eq{36b}, \eq{37}, and \eq{38} it follows that:
\be\n{39}
\mathtt{k}^{\alpha}\psi_{;\alpha}=\frac{\varepsilon}{4\omega}\varepsilon_{\alpha\beta\gamma\delta}u^{\alpha}\mathtt{k}^{\beta}\Phi^{\gamma\delta}\,.
\ee
Thus, as we already noted, the change of the polarization phase along the modified light ray is of order $\varepsilon$. This implies that the internal degree of freedom, the order $\varepsilon^0$ polarization phase $\varphi$, was indeed included into the combined eikonal.

Using the combined eikonal \eq{25a}, we can compute phase velocity of each polarization component defined with respect to its geometric optics null ray,
\be\n{ph}
v_{\text{ph}}\equiv-\frac{k_{\alpha}u^{\alpha}}{k_{\alpha}n^{\alpha}}= \frac{\omega}{\omega-\varepsilon\sigma\omega_{\alpha}n^{\alpha}}\,.
\ee 
This expression implies that the phase velocity of left- and right-handed circularly polarised light is different. This difference results in the gravitational Faraday effect discussed in Sec.~III. 

Let us now consider the dynamical equation for light rays \eq{31}. Its right-hand side depends on light polarization $\sigma=\pm1$ and represents the force of interaction between the light polarisation and space-time angular momentum. This force is orthogonal to the wave vector $\mathtt{k}^{\alpha}$. This type of polarization-dependent force is a manifestation of the gravitational spin-Hall effect of light: the back reaction of the changing polarization phase onto light trajectory results in its transverse, polarization-dependent displacement. As a result of such displacement, a linearly polarized beam of light propagating in a gravitational field of non-zero angular momentum along the space-time null geodesic will split into components of left- and right-hand circular polarizations and each component will propagate along a different null world line. These world lines get gradually displaced away in the opposite direction from the null geodesics. The group velocity of these components, which is the locally measured Poynting vector divided by the electromagnetic energy density, is equal to speed of light in vacuum. According to the generalized Fermat's principle \cite{Perlick1,Perlick2,Fermat,Frolov}, the expression \eq{48} in the next Section, integrated along a light ray trajectory, is stationary for null geodesics with respect to null variations. Thus, the propagation time $t$ for these components is greater than that of the corresponding null geodesic.       

In the next section we shall discuss the gravitational spin-Hall effect in local space and time decomposition. To conclude, we recall that source-free Maxwell equations in a 4-dimensional space-time are conformally invariant. Accordingly, the expressions above are invariant with respect to the conformal transformation of the metric $g_{\alpha\beta}=\Omega^{2}\bar{g}_{\alpha\beta}$ accompanied by the following conformal transformations:
\ba
&&\mathtt{k}^{\alpha}=\Omega^{-2}\bar{\mathtt{k}}^{\alpha}\hhh u^{\alpha}=\Omega^{-1}\bar{u}^{\alpha}\,,\n{40a}\\
\tilde{a}^2&=&\Omega^{-2}\bar{\tilde{a}}^2\hhh\mathtt{f}^{\alpha}=\Omega^{-1}\bar{\mathtt{f}}^{\alpha}+\kappa\Omega^{-2}\bar{\mathtt{k}}^{\alpha}\,,\n{40b}
\ea
where the scalar function $\kappa$ solves equation
\be\n{41}
\bar{\mathtt{k}}^{\alpha}\kappa_{,\alpha}+i\kappa\bar{\mathtt{k}}^{\alpha}\tilde{\varphi}_{;\alpha}+\bar{\mathtt{f}}^{\alpha}\Omega_{;\alpha}=0\,,
\ee
while $\omega_{\alpha}$ and $\Phi_{\alpha\beta}$ are conformally invariant, and the local angular frequency of light transforms as $\omega=\Omega^{-1}\bar{\omega}$.  

\section{Local space and time decomposition}

Let us now present the dynamical equation \eq{31} in a local decomposition defined by a static observer's 4-velocity $u^{\alpha}$. To simplify our computations, we will work in the metric $\bar{g}_{\alpha\beta}$ conformally related to the space-time metric \eq{12} via the conformal factor $\Omega^{2}=h$. In what follows, we shall drop the bar signs. Because of the conformal invariance of the dynamical equation, our final expressions will be valid in the original space-time metric $g_{\alpha\beta}$, assuming that all dynamical quantities are transformed accordingly, as shown in the end of the previous section. 

To compute the expression \eq{30} we need the covariant form of the vorticity vector (see \eq{14} and the text below),
\be\n{42}
\omega_{\alpha}=\delta^{i}_{\alpha}A_{i}\hh \BM{A}=\tfrac{1}{2}\left([\BM{g}\times\BM{g}^{\flat}_{,t}]+\text{curl}\,\BM{g}\right)\,.
\ee
Now we can calculate $\Phi_{\alpha\beta}$,
\be\n{43}
\Phi_{0i}=-{\cal E}_{i}\hhh \Phi_{ij}=e_{ijk}{\cal B}^{k}\,,
\ee
where we defined
\be\n{44}
\BM{\cal E}\equiv-\BM{A}_{,t}\hh \BM{\cal B}\equiv\text{curl}\BM{A}\,.
\ee
The 3-dimensional fields $\BM{\cal E}$ and $\BM{\cal B}$ can be considered as ``electric" and ``magnetic" components of $\Phi_{\alpha\beta}$. 

The next step is to apply the projection operators $-u^{\alpha}u_{\beta}$ and $p^{\alpha}_{\beta}=\delta^{\alpha}_{\beta}+u^{\alpha}u_{\beta}$ to the dynamical equation. These operators project 4-dimensional objects onto  $u^{\alpha}$ and its local orthogonal 3-dimensional hypersurface $\Sigma_{u}$, that gives their local space and time decomposition. However, it is more convenient to use the decomposition of the wave vector $\mathtt{k}^{\alpha}=\omega(u^{\alpha}+\mathtt{n}^{\alpha})$ [cf. \eq{4a}], where the light ray frequency $\omega=-\mathtt{k}^{\alpha}u_{\alpha}$ is measured by a local static observer and the unit space-like vector $\mathtt{n}^{\alpha}$, orthogonal to $u^{\alpha}$, defines spatial direction of the light ray. Then, we contract the expression \eq{29} with $u^{\beta}$ and $\mathtt{n}^{\beta}$ separately and apply the projection operators. Contraction with $u^{\beta}$ gives us an expression lying entirely in $\Sigma_{u}$ and involving time derivatives of the frequency and the unit space-like vector. Contraction with $\mathtt{n}^{\beta}$ followed by projection onto $u^{\alpha}$ gives us (naturally) the same expression contracted with $\mathtt{n}^{\alpha}$,
\be\n{45}
(1+g_{i}\mathtt{n}^{i})\omega_{,t}+\mathtt{n}^{i}\omega_{,i}+\frac{\omega}{2}(2\mathtt{n}^{i}g_{i,t}+\mathtt{n}^{i}\mathtt{n}^{j}\gamma_{ij,t})=\varepsilon\sigma{\cal E}_{i}\mathtt{n}^{i}\,.
\ee
Finally, projecting the expression \eq{29} contracted with $\mathtt{n}^{\beta}$ onto $\Sigma_{u}$ and using a local triad $e_{a}^{\alpha}$ defined on $\Sigma_{u}$ gives us the light ray dynamical equation on $\Sigma_{u}$,
\ba
&&\hspace{-0.5cm}\mathtt{n}^{b}\mathtt{n}^{a}_{\,\,\,|b}=\frac{\omega_{,b}}{\omega}\Pi^{ab}+2\left[\BM{\mathtt{n}}\times\BM{A}\right]^{a}+\frac{\varepsilon\sigma}{\omega}\left[\BM{\mathtt{n}}\times\BM{\cal B}_{g}\right]^{a}\,,\n{46}\\
&&\hspace{2cm}\BM{\cal B}_{g}\equiv\BM{\cal B}+[\BM{\cal E}\times\BM{g}]\,.\n{47}
\ea
Here $\mathtt{n}^{b}\mathtt{n}^{a}_{\,\,\,|b}$ is the covariant derivative defined in the metric $\gamma_{ab}=e_a^{i}e_b^{j}\gamma_{ij}$ along a light ray trajectory with the unit tangent vector $\mathtt{n}^{\alpha}=dx^{\alpha}/dl=e_{a}^{\alpha}\mathtt{n}^{a}$, where $dl=\omega d\lambda$ is the proper distance in the 3-dimensional hyperspace $\Sigma_{u}$, and $\Pi^{ab}=\gamma^{ab}-\mathtt{n}^{a}\mathtt{n}^{b}$ is the 3-dimensional projection operator onto a local 2-dimensional subspace orthogonal to $\mathtt{n}^{a}$. The proper distance $dl$ is related to the time coordinate $t$ as follows: 
\be\n{48}
\frac{dt}{dl}=1+(\BM{g}, \BM{n})\,,
\ee
where the scalar product is defined in the metric $\gamma_{ab}$. The expressions \eq{45}, \eq{46}, and \eq{48} represent the local space and time decomposition of the dynamical equation \eq{31}. 

Let us now analyse the derived result. Our main object is the dynamical ray equation \eq{46}. To understand better its physical meaning, we introduce the optical metric $\pi_{ab}$ and the 3-dimensional wave vector $\Bbbk_{a}$ as follows:
\be\n{49}
\pi_{ab}\equiv\omega^{2}\gamma_{ab}\hhh \Bbbk_a\equiv\omega\mathtt{n}_{a}\,.
\ee
Then, the dynamical ray equation \eq{46} takes the following form:
\be\n{50}
\frac{D\BM{\Bbbk}}{d\ell}=2\left[\BM{\Bbbk}\times\BM{A}\right]+\frac{\varepsilon\sigma}{\omega}\left[\Bbbk\times\BM{\cal B}_{g}\right]\,.
\ee
Here $D\Bbbk/d\ell$ is the covariant derivative of the wave vector defined in the optical metric, $d\ell=\omega dl$ is the optical length, and the vector products are defined in the optical metric. This equation is similar to the dynamical equation of a charged particle moving in a non-inertial rotating frame in the presence of a magnetic field, i.e. the term $2\left[\BM{n}\times\BM{A}\right]$ is the Coriolis force and the next term is the Lorentz force. This last term is the transverse, polarization-dependent force that gives rise to the gravitational spin-Hall effect of light. This force depends on light frequency. Thus, the resultant splitting of a non-monochromatic beam of light onto left- and right-handed circularly polarized components is frequency dependent. Namely, the low frequency part of the beam gets more deflected in the transversal direction, as compared to the high frequency part.  As a result of such deflection, we have a {\em polarized gravitational rainbow}.

To conclude, we remark that in the case of a stationary space-time, our static observers are Killing observers and the results above reduce to those derived in \cite{Frolov:2011mh}.
 
 \section{Discussion}
 
Here we presented the study of the gravitational Faraday and spin-Hall effects of light in arbitrary, non-stationary, asymptotically flat space-of non-zero angular momentum. This is a generalisation of the modified geometric optics formalism developed in \cite{Frolov:2011mh} and applied to propagation of polarized light in the stationary space-time of a rotating (Kerr) black hole \cite{Frolov:2012zn}. The key concept of this generalisation is the field of static observers defined in Sec.~IV.  In a stationary space-time, this field of observers naturally reduces to the field of Killing observers. The field of observers is uniquely defined in accordance with vanishing space-time ADM 3-momentum. The gravitational Faraday and spin-Hall effects are described properly by such static observers. One may try to explore the formalism developed here in the language of Finsler geometry and the related Randers metric applied to sationary spacetimes in \cite{Gibbons:2008zi}.

The natural limitation of our formalism is breakdown of the geometric optics applicability and the concept of the static observers field. As we already pointed out in Sec.~IV (see footnote 4), such observers cannot exist in space-time regions where a rigid static latticework spreading out of asymptotically flat region cannot be defined, for example in ergoregions or black hole interior. Also, light emitted form the vicinity of the space-time ergoregions or form the vicinity of a static black hole horizon and propagating to the asymptotically flat region of the space-time gets strongly redshifted, such that its frequency can take very small values. This results in breakdown of the geometric optics approach. Let us also note that detection of the gravitational Faraday and spin-Hall effects of light requires measurements of a very high sensitivity. These effects are due to strong non-static gravitational fields and they accumulate during light propagation. The angular split of left- and right-handed circularly polarized light components is proportional to the space-time angular momentum \cite{Mashhoon1,Mashhoon2,Mashhoon3,Mashhoon4,Frolov:2012zn}, whereas spatial separation of these components is an accumulative effect and proportional to the propagation distance. Thus, the effect may not easily be observable in weak gravitational fields and relatively small spatial regions, e.g., within the Solar System \cite{Mashhoon3}, but it could potentially be detected in the light emerged from strongly gravitating systems, such as vicinity of a black hole, and propagated sufficiently large distance. Finally, we remark that the approach developed here can be adapted to describe gravitational Faraday and spin-Hall effects of weak gravitational waves.

\end{document}